# Learning the aerodynamic design of supercritical airfoils through deep reinforcement learning

Runze Li,* Yufei Zhang,† Haixin Chen‡

(*Tsinghua University, Beijing, 100084, People's Republic of China*)

**The aerodynamic design of modern civil aircraft requires a true sense of intelligence since it requires a good understanding of transonic aerodynamics and sufficient experience. Reinforcement learning is an artificial general intelligence that can learn sophisticated skills by trial-and-error, rather than simply extracting features or making predictions from data. The present paper utilizes a deep reinforcement learning algorithm to learn the policy for reducing the aerodynamic drag of supercritical airfoils. The policy is designed to take actions based on features of the wall Mach number distribution so that the learned policy can be more general. The initial policy for reinforcement learning is pretrained through imitation learning, and the result is compared with randomly generated initial policies. The policy is then trained in environments based on surrogate models, of which the mean drag reduction of 200 airfoils can be effectively improved by reinforcement learning. The policy is also tested by multiple airfoils in different flow conditions using computational fluid dynamics calculations. The results show that the policy is effective in both the training condition and other similar conditions, and the policy can be applied repeatedly to achieve greater drag reduction.**

* Ph. D. student, School of Aerospace Engineering, email: lirz16@mails.tsinghua.edu.cn

†Associate professor, School of Aerospace Engineering, senior member AIAA, email: zhangyufei@tsinghua.edu.cn

‡ Professor, School of Aerospace Engineering, associate fellow AIAA, email: chenhaixin@tsinghua.edu.cn

## Nomenclature

| | | |
|---|---|---|
| $AoA$ | = | Angle of attack |
| $\boldsymbol{a}$ | = | Action |
| $C_L$ | = | Lift coefficient |
| $C_D$ | = | Drag coefficient |
| $h_b$ | = | Height of the bump |
| $M_w$ | = | Wall Mach number |
| $M_\infty$ | = | Free-stream Mach number |
| $Q$ | = | State-action value function |
| $r$ | = | Reward |
| $R$ | = | Cumulative reward |
| $Re$ | = | Reynolds number |
| $\boldsymbol{s}$ | = | State |
| $s_b$ | = | Width of the bump |
| $t_{max}$ | = | Maximum airfoil thickness |
| $t_1$ | = | Center location of the bump |
| $t$ | = | Step |
| $T$ | = | Length of a trajectory |
| $V$ | = | State value function |
| $X$ | = | Location |
| $\boldsymbol{\pi}$ | = | Policy |
| $\boldsymbol{\phi}$ | = | Value function parameters |
| $\tau$ | = | Trajectory |
| $\boldsymbol{\theta}$ | = | Policy parameters |

# I. Introduction

With the development of computational fluid dynamics (CFD), computer-aided engineering based on large-scale calculations and analyses has become increasingly important in the field of aircraft design. After decades of aerodynamic design by trial-and-error, shape optimization has been studied and applied since the 1960s. In recent years, artificial intelligence (AI) technologies have attracted attention since they can better utilize data and improve optimization efficiency. The surrogate model is one of the first and most popular technologies applied in aerodynamic research and optimization [1]. Numerous surrogate models have been studied to make predictions based on CFD data [2-4]. With the development of artificial neural networks (ANNs), many AI methods based on ANNs or deep neural networks (DNNs) have been studied to improve the efficiency of aerodynamic optimization [5,6], such as convolutional neural networks (CNNs) [7,8], and generative adversarial networks (GANs) [9,10]. Most of these technologies take advantage of AI in extracting features, memorizing data, or making predictions.

However, AI ultimately lightens the load on people, so it is expected to be intelligent and able to mimic the behavior of people. In the field of aerodynamic design, although AI and optimization methods can accelerate the process of achieving the desired objective, the optimized results may require further modification by designers. Because there are always tradeoffs between considerations that are difficult to mathematically describe in optimizations, it is the extensive knowledge and rich experience of designers that should be learned by AI. There have been several studies that extract aerodynamic knowledge based on statistical analysis [11-13]; however, their applications are limited to improving optimizations in specific conditions, and general experience and skills are still not acquired.

In contrast, reinforcement learning is artificial general intelligence, which has the potential of learning how to do things in the same way as humans [14,15]. It can learn skills and accumulate experiences by autonomously interacting with environments without additional human instructions or supervision. Gaming is the most popular application area, in which reinforcement learning can achieve superhuman performance in numerous games, such as Go board games and Atari games [16,17]. In robotics, reinforcement learning can provide robots with the ability to learn tasks and even adapt a learned skill to a new task, which is called the ability of transfer application [18]. Recently, reinforcement learning achieved a breakthrough in electronic chip design [19], which first showed that it can be used in an actual industrial design. Therefore, these developments indicate that reinforcement learning can develop the skills of aircraft aerodynamic design and achieve the ability of transfer application.

The present paper proposes a physical perspective of utilizing reinforcement learning in the aerodynamic design of transonic supercritical airfoils. Shape optimizations search for optimized airfoils based on geometries and performances. Because the airfoil characteristic nonlinearly changes when the flow condition is changed, most surrogate models in AI applications are only valid for the training condition. In contrast, human designers make geometry modifications based on their observations of geometries, performances, pressure distributions, and even details of the flow field. Then, their experiences and skills are more general and usually can be applied in similar conditions. Therefore, the present paper learns the design policies of modifying airfoils based on performances and pressure distributions. Consequently, reinforcement learning can formulate general and physical policies that can be applied in other situations.

The present paper utilizes the proximal policy optimization algorithm (PPO) to learn the policy of supercritical airfoil drag reduction. After introducing the background and key elements of the reinforcement learning method, the paper is organized as the policy learning procedure shown in Figure 1. First, surrogate models are built based on airfoil samples evaluated by CFD so that the computational cost of policy learning can be reduced. Second, the implementation of the environment for airfoil drag reduction is discussed. Third, to improve the efficiency of reinforcement learning, the initial policy is pretrained by imitation learning of good state-action samples. Then, the drag reduction policy based on the airfoil pressure distribution is trained by the PPO algorithm. Finally, the policy is tested by CFD in other situations to indicate the transfer application ability. Additionally, the learned policy based on the physical features of the pressure distributions is compared with a comparable policy based on airfoil geometries. Both policies are tested by CFD in a test flow condition, and the average drag reductions are compared.

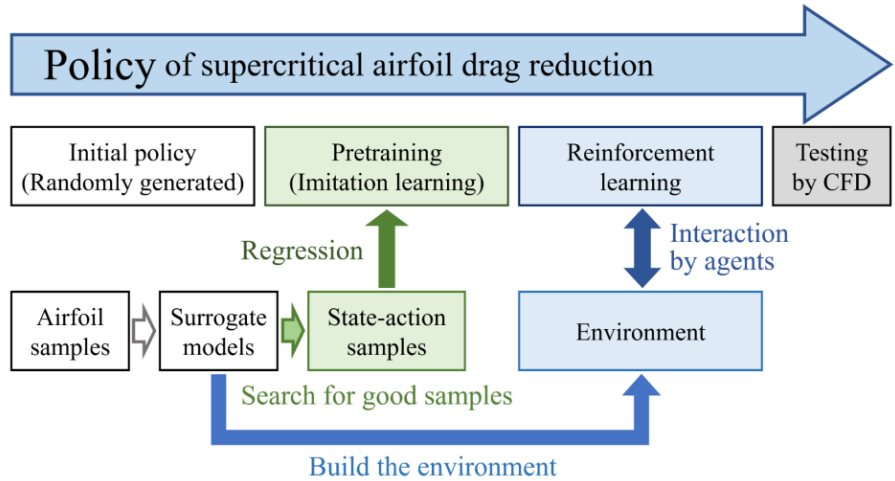

**Figure 1 Airfoil design policy learning procedure**

## II. Reinforcement learning for airfoil aerodynamic design

Reinforcement learning is a machine learning technology that mimics the human learning process. Different from directly approximating functions in supervised learning, reinforcement learning does not directly theorize or approximate how people make decisions. There are a limited number of studies of reinforcement learning in the field of fluid dynamics, most of which utilized reinforcement learning for active control problems [20,21], and very few of them attempted shape optimizations [22,23]. The present paper utilizes reinforcement learning for airfoil drag reduction and formulates its policy by interacting with the environment. Reinforcement learning is not told which action to take, but instead, it must discover which action can achieve more reward by attempting it. Therefore, the environment and many other key elements, e.g., agent, action, state, reward, policy, trajectory and value, of reinforcement learning must be properly defined for aerodynamic design. Figure 2 shows the basic process of reinforcement learning, i.e., the agent interacts with its environment, and the relationships between most of the key elements are also indicated.

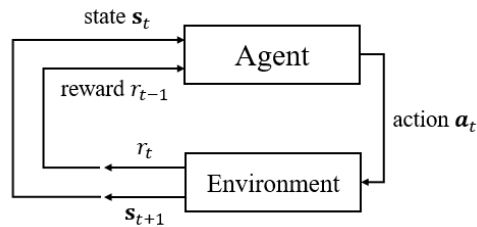

**Figure 2 Interaction between an agent and its environment**

In this section, the background of reinforcement learning is first introduced, and then the key elements are defined in relation to the airfoil drag reduction problem. Correspondingly, the CFD and geometry parameterization methods are also presented. Additionally, to reduce the computational cost of reinforcement learning, the environment utilizes fast analysis methods using ANNs as surrogate models. After setting up the airfoil sample sets and the surrogate models, algorithms for reinforcement learning and the environment are described.

**A. Reinforcement learning**

Reinforcement learning learns the policy by using agents to interact with its environment [14]. As shown in Figure 2, each agent takes an action $a_t$ based on its state $s_t$ in time step $t$. Then, the environment takes the state and action as input and returns the next state $s_{t+1}$ of the agent and the reward $r_t$ of this step. Then, as shown in Figure 3, the agent starts from its initial state $s_0$ and keeps interacting with the environment by constantly taking actions and updating its state until it meets the stop signal. The sequence of states and actions that this agent experiences is called a trajectory, i.e., $\tau = \{s_0, a_0, \cdots s_{T-1}, a_{T-1}, s_T\}$. Therefore, the goal of reinforcement learning is to formulate a policy $\pi$ for agents to take actions so that they can achieve the highest cumulative reward, i.e., $R(\tau) = \sum_\tau r_t$.

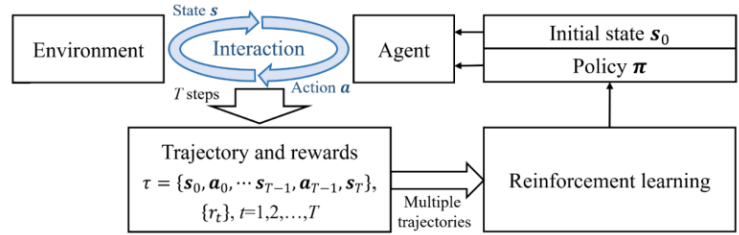

**Figure 3 Process of reinforcement learning**

Reinforcement learning can be roughly categorized into model-based methods and model-free methods [14]. Model-based methods incorporate a model of the environment, which is the function approximation of the environment [24]. However, each environment will need a different model representation, and it is essentially similar to surrogate-based optimizations. Therefore, model-based methods are not frequently discussed in reinforcement learning studies. In contrast, model-free methods directly interact with environments, and they can be further classified into policy-based and value-based methods.

Policy-based methods directly optimize the policy, whereas value-based methods take a relatively indirect path [14]. The value function, i.e., $V(s)$ or $Q(s, a)$, is a concept similar to cumulative rewards. Reward indicates what is good in an immediate sense. A value function indicates the long-term desirability of states (and actions). Roughly speaking, the value of a state, i.e., the state value function $V(s_t)$, is the cumulative reward that an agent is expected to receive in the future starting from this state $s_t$. The state value function of policy $\pi$ is usually defined as $V^\pi(s_t) = \mathop{E}_{\tau \sim \pi}[R(\tau)|s_t]$, in which the trajectory $\tau$ is taken according to the policy $\pi$. The state-action value function, i.e., Q function $Q(s_t, a_t)$, is the cumulative reward when an agent takes an action $a_t$ starting from this state $s_t$. It is usually defined as $Q^\pi(s_t, a_t) = \mathop{E}_{\tau \sim \pi}[R(\tau)|s_t, a_t]$. Then, it is straightforward to see that $V^\pi(s_t)$ is the weighted average of $Q^\pi(s_t, a_t)$ over all possible actions by the possibility of each action, i.e., $V^\pi(s_t) = \mathop{E}_{a \sim \pi}[Q^\pi(s_t, a_t)]$.

The value-based methods estimate the value function $Q(s, a)$ and then determine the action by a predetermined policy. For example, the action $a_t$ can be determined by finding the optimal action that maximizes the value function $Q(s_t, a)$ in the current state $s_t$. There is another category of methods that simultaneously exploit the policy and value function, i.e., the actor-critic (AC) methods. These methods estimate the value function, i.e., the critic, to 'criticize' the policy of its actor to improve learning performance.

The policy and value function can be stored in tables when the state and action are low-dimensional discrete values. However, for most situations that have continuous or high dimensional states and actions, the policy and value function have to be estimated by ANN. Then, these reinforcement learning algorithms using deep neural networks as function approximators are called deep reinforcement learning [15].

The policy of reinforcement learning can be either deterministic or stochastic [25]. The deterministic policies $a = \pi(s)$ always take the same action when facing the same state. The stochastic policy samples the action according to the probability distribution specified by a function of the state-action pair, i.e., $\pi(s, a) = P(a_t = a|s_t = s)$. Since reinforcement learning learns by rewarding desired actions and punishing undesired actions, both desired and undesired actions may occur in each step when using stochastic policies, which provides a clearer direction to adjust to. Therefore, the stochastic policy is more robust and easier to optimize via gradient methods. Consequently, the value function for stochastic policies has to be estimated via Monte Carlo methods.

Figure 4 shows the modules of reinforcement learning for airfoil drag reduction. Detailed implementations are introduced in the following sections. In this paper, the PPO algorithm is utilized to improve a stochastic policy $\pi$ for reducing airfoil drag. Multiple agents with different initial states are used to obtain a more general policy. Each agent is assigned a different initial airfoil from a specific airfoil sample set, and each agent takes multiple trajectories by Monte Carlo sampling of the policy. For every trajectory, the agent interacts with the environment. After receiving an action $a$, the environment utilizes the class-shape function transformation (CST) method and bump functions to modify airfoil geometries. Then, the new airfoil is evaluated by CFD or surrogate models so that the drag coefficient and pressure distribution features can be obtained. Consequently, the reward $r$ and next state $s$ are collected. All the trajectories are used to improve the current policy by the PPO algorithm; then, the new policy is employed for sampling trajectories in the next iteration of reinforcement learning.

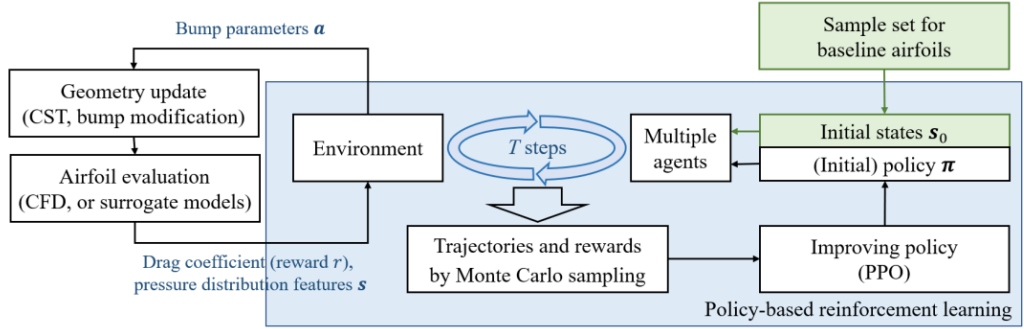

Figure 4 Implementation of reinforcement learning modules

**B. Key elements for airfoil drag reduction**

From the airfoil aerodynamic design point of view, reinforcement learning mimics the process of designer learning manual design skills by trial-and-error. The policy that reinforcement learning improves is the design experience. The agent, i.e., the software version of designers, makes observations of the current design, such as the airfoil geometry, performances, or features of the flow field. The state, which is usually the combination of these observations, is used to describe the situation in which the agent finds itself. Then, the agent employs the policy to determine an action based on the current state. Usually, the action is the geometry modification to the current airfoil. Or, to be more precise, the action vector contains the values of geometry modification parameters, e.g., the change in airfoil thickness or camber distribution, or parameters of airfoil bump modification functions. Therefore, the environment is the module that updates the airfoil geometry and evaluates the performances so that the next state and reward can be provided.

Reinforcement learning learns how to obtain a good cumulative reward over many steps. Fortunately, the drag reduction of supercritical airfoils has dense rewards, i.e., the reward can be obtained in each step, which is defined as the drag coefficient ($C_D$, measured in drag counts, 1 count equals 0.0001 of the drag coefficient) reduction, i.e., $r_t =$ $R(\boldsymbol{s}_t, \boldsymbol{a}_t) = 10{,}000 \times (C_{D,t} - C_{D,t+1})\big|_{s_t,a_t}$. This makes it a relatively easier problem for reinforcement learning than problems with delayed rewards, such as Go board games.

State and action are the other two key elements that need to be carefully defined since they ultimately determine the performances of reinforcement learning and the learned policy. Because the performances of an airfoil are comprehensively associated with its flow field, the policy should imitate general design experiences based on aerodynamics and features of flow fields so that the policy can be applied in different situations. Statistical features of flow fields have been extracted by machine learning methods and used for drag predictions [26] or reinforcement learning of control problems [27]. Physical features of flow fields, especially features of pressure distribution on the airfoil surfaces, have been utilized for optimizations and surrogate models as well. They can capture physical laws with fewer parameters and achieve better interpretability for researchers and designers [4,28]. In this paper, the physical features of pressure distribution on airfoil surfaces are chosen as the state parameters. Since the Mach number is more consistent for describing the physical relationship, the pressure distribution is further represented by the wall Mach number distribution. The wall Mach number ($M_w$) is the Mach number calculated based on an isentropic relationship with the pressure coefficient on the airfoil surface and free-stream Mach number $M_\infty$ [29].

Several features of wall Mach number distributions are discussed in the present paper and are illustrated in Figure 5. Detailed definitions can be seen in [28].

(1) Wall Mach number of the suction peak, $M_{wL}$.

(2) Location of the shock wave, $X_1$.

(3) Wall Mach number in front of the shock wave, $M_{w1}$.

(4) Highest wall Mach number on the lower surface, $M_{w,lower}$.

(5) Smoothness of the suction plateau, $Err$.

(6) Highest wall Mach number behind the shock wave, $M_{w,A}$. This parameter is defined to describe the secondary flow acceleration behind the shock wave that sometimes may occur. When there is no secondary flow acceleration, as shown in Figure 5, $M_{w,A}$ is $M_w$ just behind the shock wave.

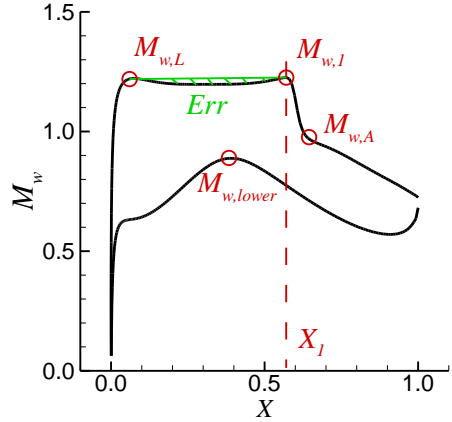

**Figure 5 Definitions of wall Mach number distribution features**

Because the airfoil drag coefficient mostly depends on the wave drag on the upper surface, only the airfoil upper surface is modified in the environment to simplify the problem. Therefore, the features on the upper surface are more important for taking actions, and the state of the reinforcement learning is thus defined as a vector of four elements, i.e., $s = [X_1, M_{w,1}, M_{w,L}, M_{w,A}]$. The action $a$ is defined as a vector containing parameters for geometry modification of the airfoil upper surface, which is discussed in the following section.

## C. Geometry parameterization and CFD methods

There are two geometry parameterization methods in this reinforcement learning problem: the CST method [30], which represents the airfoil geometry by combining several polynomials, and the Hicks-Henne bump function [31], which is used for the local modification of the airfoil upper surface. The CST method can describe an arbitrary geometry and guarantee smoothness with comparatively fewer parameters. A sixth-order Bernstein polynomial is used as the primary function of the CST method; i.e., seven CST parameters are used to describe upper and lower surfaces. The Hicks-Henne bump function can achieve local geometry modifications while keeping the surface smooth. The bump modification method adds incremental bump functions $y_{bump}$ to the baseline curve $y_{baseline}$, as described in Eq. 1, in which the Hicks-Henne bump function can be described as Eq. 2. $t_1 \in (0,1)$ is the location of the highest point of the bump function, and $|h_b|$ is the maximum height. $t_2$ controls the bump width $s_b$, and $s_b$ is defined as the length between the two points on both sides of $t_1$, which have a height of $0.01h_b$. Then, $t_2$ can be calculated once the bump width is specified. Additionally, the $\pi$ in Eq. 2 is the mathematical constant, whereas other bold $\boldsymbol{\pi}$ in the present paper represents the policy.

$$y = y_{baseline} + \sum_i y_{bump,i} \tag{1}$$

$$y_{bump} = h_b \left[\sin\left(\pi x^{\frac{\log 0.5}{\log t_1}}\right)\right]^{t_2}, x \in [0,1] \tag{2}$$

Figure 6 shows some examples of the Hicks-Henne bump function, which are defined by the three parameters. Therefore, the action vector can be defined as $\boldsymbol{a} = [t_1, s_b, h_b]$. The ranges of the three components are listed in Eq. 3.

$$t_1 \in (0,1), s_b \in [0.2, 0.4], h_b \in [-0.1, 0.1] \tag{3}$$

Figure 7 shows the process of taking an action $\boldsymbol{a} = [0.3, 0.4, 0.02]$, i.e., adding the green curve in Figure 6 to the baseline airfoil upper surface (black dashed curve) in Figure 7. The green solid curve in Figure 7 shows the geometry and curvature after adding the incremental bump function. Although the bump function itself is smooth, it causes a large fluctuation in the curvature distribution. Therefore, the surface is further smoothed to obtain a reasonable airfoil, i.e., the blue solid curve in Figure 7. The smoothed curve is obtained by using the 6th-order CST method to reconstruct the modified airfoil so that the curvature distribution is more reasonable, and the CST parameters representing the airfoil are also updated. The airfoil maximum relative thickness $t_{max}$ is kept at 0.095 during the modification process. Therefore, if the maximum thickness changes after the bump modification, the airfoil lower surface will be scaled to maintain the maximum relative thickness.

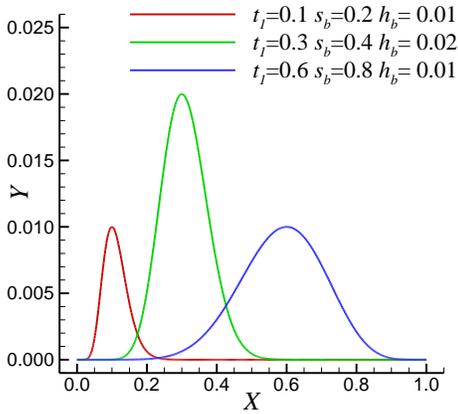

**Figure 6 Examples of bump functions**

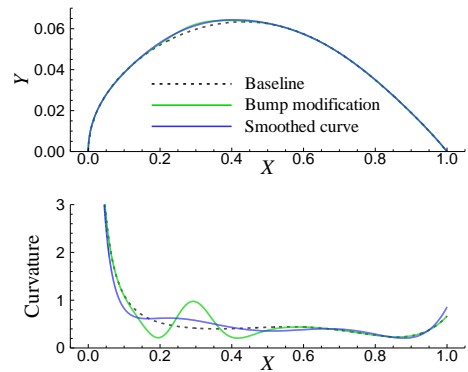

**Figure 7 Changes in geometry and curvature during the bump modification**

After constructing the airfoil geometry, a C-grid is generated for Reynolds-averaged Navier–Stokes (RANS) analysis. The CFD solver is a well-known open source solver named CFL3D [32]. In this paper, the monotonic upwind scheme for conservation laws (MUSCL), Roe's scheme, the lower-upper alternating direction implicit algorithm (LU-ADI), and the shear stress transport (SST) model are selected for reconstruction, spatial discretization, time advancement, and turbulence modeling, respectively. Figure 8(a) shows the experimental pressure coefficient ($C_p$) distributions of an RAE2822 airfoil [33] compared with the CFD results of different grid sizes. The free-stream Mach number ($M_\infty$) is 0.725, the Reynolds number ($Re$) based on the unit chord length is $6.5 \times 10^6$, and the angle of attack ($AoA$) is 2.55 degrees. The three grids, with sizes of 20,000, 40,000, and 80,000, respectively. The Δy+ of the first grid layer is always kept less than 1. All grids can achieve a similar $C_p$ resolution. The medium grid is shown in Figure 8(b), which is employed in this paper.

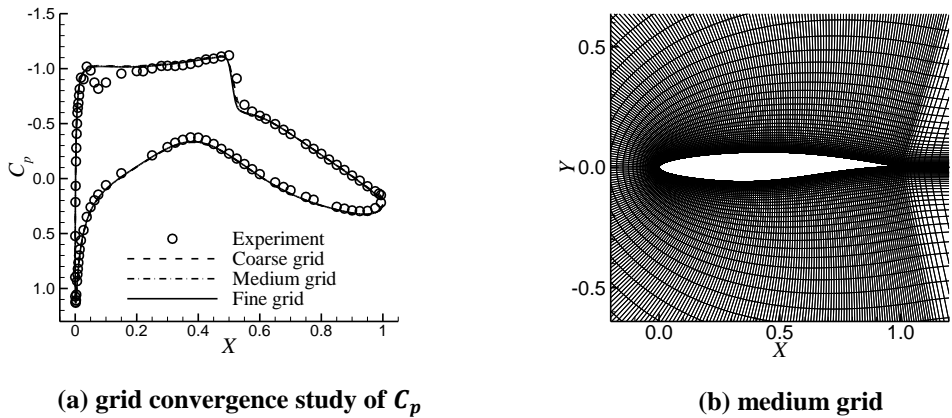

(a) grid convergence study of $C_p$  (b) medium grid

**Figure 8 CFD validation and grid**

The flow condition is that the free-stream Mach number ($M_\infty$) is 0.76, the Reynolds number based on the unit chord length ($Re$) is $5 \times 10^6$. The lift coefficient ($C_L$) is kept at 0.70 for all airfoils, by adaptively adjusting the angle of attack during the RANS simulation to make sure that it converges to the specified lift coefficient.

**D. Fast analysis method and sample sets**

Reinforcement learning can operate in any situation as long as a clear reward is applied. However, reinforcement learning is sometimes difficult to deploy due to its reliance on exploration of the environment. When being conducted on an expensive environment, such as the environment relaying CFD computations, reinforcement learning will cause unbearable costs because of its constantly seeking new states and attempting different actions. Therefore, fast analysis methods are necessary for conducting reinforcement learning. Since the state is based on the wall Mach number distribution features and the reward is the reduction in drag coefficient, surrogate models for $C_D$, $X_1$, $M_{w,1}$, $M_{w,L}$ and $M_{w,A}$ are needed for construction.

The airfoil sets for training and testing are obtained by an adaptive sampling method based on radial basis function (RBF) response surfaces [34]. Adaptive sampling is designed to obtain samples with various geometries and wall Mach number distribution features. Because airfoils with a single shock wave and reasonable $C_D$ are an interest in industrial design, only the samples that have their features within the given range specified in Eq. 4 are sampled.

$$C_D \in [0.009, 0.013], X_1 \in [0.2, 0.8]$$
$$M_{w,1} \in [1.0, 1.2], M_{w,L} \in [1.0, 1.3], M_{w,A} \in [0.9, 1.1]$$
(4)

The adaptive sampling increases the diversity of the airfoil shock wave location and suction peak, while the Mach number before the shock wave, i.e., $M_{w,1}$, keeps approximately three values, 1.12, 1.14, and 1.16. A total of approximately 10,000 airfoils are generated. A total of 5,000 airfoils are selected for the training set for building the surrogate models, and 200 airfoils are selected for testing. The selection process is described by Algorithm A so that all the airfoils have features within the given range in Eq. 4, and the geometries are as different from each other as possible. Figure 9 shows the selected airfoils (gray curves), and the colored curve 1, 2, and 3 are some typical wall Mach number distributions of the selected airfoils. It shows that the sample sets cover all concerned possibilities, and they are adequate for building the surrogate models.

Algorithm A: sample selection

1: Eliminate all the samples that do not satisfy the constraint in Eq. 4;
2: While (sample number > 200);
3:   Calculate Euclidean distances of CST parameters between all samples;
4:   Delete the sample with the smallest distance;
5:   If the total number of samples equals 5,000, save the current samples to the training set;
6:   If the total number of samples equals 200, save the current samples to the testing set;
7: end

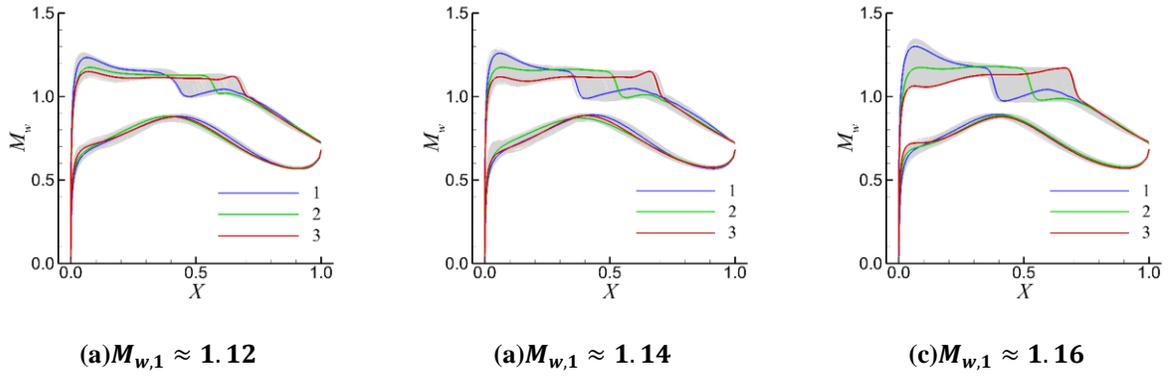

(a) $M_{w,1} \approx 1.12$      (a) $M_{w,1} \approx 1.14$      (c) $M_{w,1} \approx 1.16$

**Figure 9 Valid airfoil samples and typical samples [34]**

The ANN for surrogate models is constructed using PyTorch [35]. There are 3 hidden layers and 1,024 neurons in each layer. The input layer has 14 neurons representing the CST parameters, and the output layer size is five, i.e., $C_D$, $X_1$, $M_{w,1}$, $M_{w,L}$ and $M_{w,A}$. The ANN is trained by a minibatch gradient descent algorithm [36], in which the minibatch size is 128. The learning rate is 0.01 for 200 epochs, 0.001 for the next 200 epochs, 0.0001 for the next 400 epochs, and 0.00001 for the last 400 epochs. Figure 10 shows the prediction errors of drag coefficients and the wall Mach number distribution feature, and the results on both the training set (solid lines) and the testing set (dashed lines) are included. The prediction error is the relative root mean square error (RSME) defined as Eq. 5, of which the upper and lower bounds are the bounds defined in Eq. 4. The prediction error is plotted for every 100 minibatches in Figure 10. The results show that most objectives can achieve a relatively good prediction accuracy.

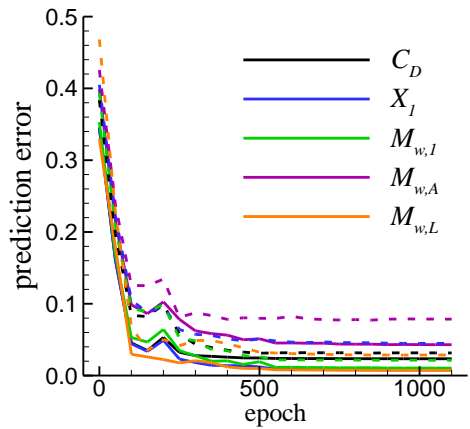

**Figure 10 History of prediction errors on the training and testing sets**

$$RSME(x) = \frac{1}{x_{upp} - x_{low}} \left[ \sum_{i=1}^{n} \frac{(x_i - \bar{x})^2}{n} \right]^{1/2} \tag{5}$$

There are another two sample sets utilized in the present paper. All the sample sets are summarized in Figure 11. Sample set No. 1 is the 5,000 airfoils for training the surrogate models, and sample set No. 2 is the testing set for the surrogate models. Sample set No. 3 is a subset with 50 airfoils randomly selected from sample set No. 2. Both sets No. 2 and 3 are used for policy training. Another sample set (No. 4) is selected from sample set No. 1 for testing the policies, it has 35 airfoils with typical wall Mach number distributions. The wall Mach number distributions of the 35 airfoils are shown in Figure 12, the average drag of airfoil No. 1-10, No. 11-24, No. 25-35 is 101.56, 103.64, 107.19 drag counts.

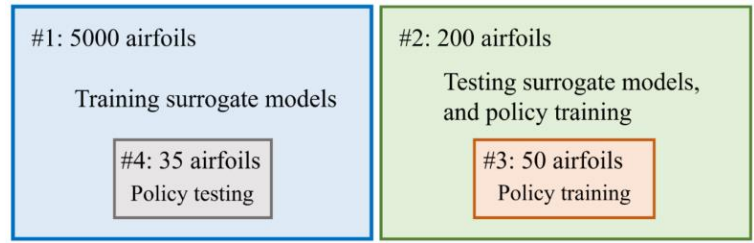

Figure 11 Sample sets for training and testing

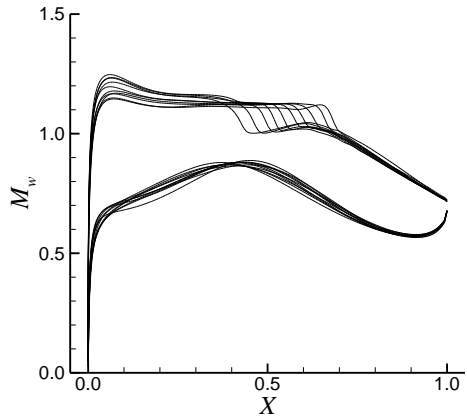

(a) airfoil No. 1-10 ($M_{w,1} \approx 1.12$)

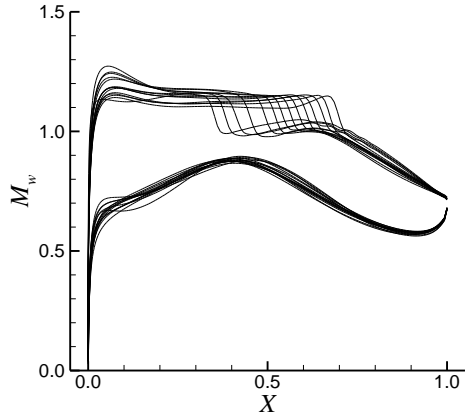

**(b) airfoil No. 11-24 ($M_{w,1} \approx 1.14$)**

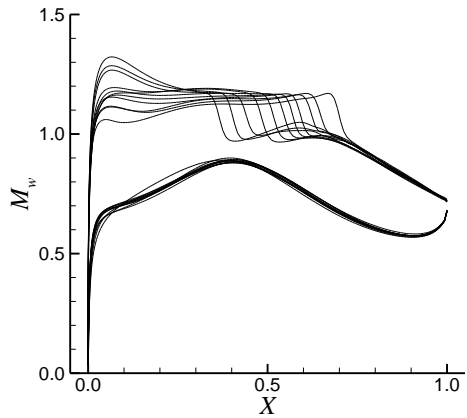

**(c) airfoil No. 25-35 ($M_{w,1} \approx 1.16$)**

**Figure 12 Wall Mach number distributions of 35 typical airfoils**

### E. Proximal policy optimization and environment

Proximal policy optimization (PPO) is a popular reinforcement learning algorithm that has been successfully applied in many situations [37]. PPO is a deep reinforcement learning algorithm categorized as a policy gradient method. PPO has some of the benefits of trust region policy optimization (TRPO) algorithm [38], but it is much simpler to implement, more general, and have better sample complexity. It can also utilize the generalized advantage estimation (GAE) [39] to improve its efficiency.

The PPO algorithm utilizes an actor-critic structure. As shown in Figure 13, the stochastic policy is represented by an actor ANN and a standard deviation layer, and the value function is the critic ANN. The size of each layer is listed in parentheses. Both the actor and critic ANNs have two hidden layers with 512 neurons, and the size of both input layers is four, i.e., the dimension of state $s$. The size of the actor output layer is three, i.e., the dimension of action $a$, and the critic output layer is the scaler value function $V_\phi$. Since the stochastic policy is employed in PPO, when a state is given, the action $a$ is sampled from the multivariate Gaussian probability distribution parameterized by a mean value $\boldsymbol{\mu}_a$ and a standard deviation $\mathbf{std}_a$. The actor ANN outputs mean values of actions. The standard deviation of actions is outputted by the standard deviation layer, which is a single layer of three neurons storing the value of standard deviations. Then, the policy and value function are updated by the most commonly used PPO algorithm by interacting with the environment [37], i.e., the PPO-clip, which is described by Algorithm B and Figure 14.

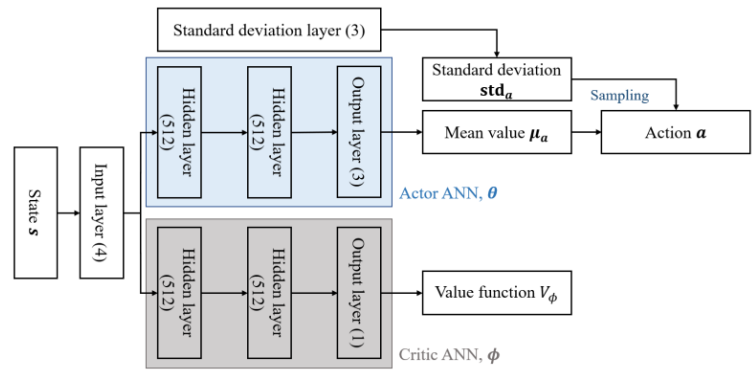

**Figure 13 Agent structure**

Algorithm B: PPO-clip

1: Initialize the policy (actor) parameters $\boldsymbol{\theta}_0$ and the value function (critic) parameters $\boldsymbol{\phi}_0$;

2: for k = 0, 1, 2, … (PPO iterations);

3:     Get $n_\tau$ trajectories $D_k = \{\tau_i\}$ based on the current policy $\pi_{\boldsymbol{\theta}_k}$;

4:     Compute the reward-to-go $\hat{R}_t$ of each step $t$ in each trajectory,
where $\hat{R}_t = \frac{1}{n_\tau T}\sum_{\tau \in D_k}\sum_{t'=t}^{T} r_{t'}$;

5:     Compute the advantage estimates $\hat{A}^{\pi_{\theta_k}}$ based on the current value function model $V_{\boldsymbol{\phi}_k}$, using the GAE algorithm [39];

6:     Update the actor parameters $\boldsymbol{\theta}$ and standard deviation $\mathbf{std}_a$ by stochastic gradient descent algorithms, e.g., Adam. The loss function is $\mathrm{loss}_{\mathrm{actor}} - 0.001\mathrm{loss}_{\mathrm{entropy}}$:

$$\mathrm{loss}_{\mathrm{actor}} = -\frac{1}{n_\tau T}\sum_{\tau \in D_k}\sum_{t=0}^{T} \min\left[\frac{\pi_\theta(s_t,a_t)}{\pi_{\theta_k}(s_t,a_t)}\hat{A}^{\pi_{\theta_k}}(s_t,a_t),\ g\left(\epsilon, \hat{A}^{\pi_{\theta_k}}(s_t,a_t)\right)\right],$$

$$\mathrm{loss}_{\mathrm{entropy}} = 0.5 + 0.5\log(2\pi) + \log\left(\prod_i \mathrm{std}_{a,i}\right),$$

where $g(\epsilon, A) = \begin{cases}(1+\epsilon)A, & A \geq 0 \\ (1-\epsilon)A, & A < 0\end{cases}$, $\epsilon = 0.1{\sim}0.3$, $\mathrm{std}_{a,i}$ is the $i^{\mathrm{th}}$ component of $\mathbf{std}_a$ vector;

7:     Fit the critic parameters $\boldsymbol{\phi}$ by regression using stochastic gradient descent algorithms:

$$\mathrm{loss}_{\mathrm{critic}} = \frac{1}{n_\tau T}\sum_{\tau \in D_k}\sum_{t=0}^{T}\left[V_{\boldsymbol{\phi}_k}(s_t) - \hat{R}_t\right]^2;$$

8: end for

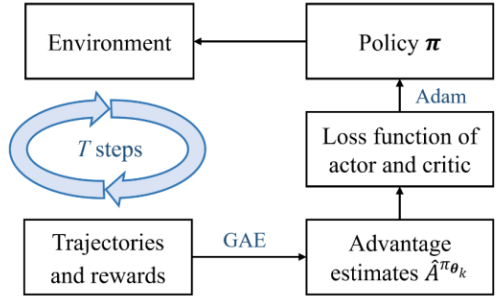

**Figure 14 PPO implementation**

In this paper, the number of trajectories in one PPO iteration is $n_\tau$. When 50 baseline airfoils in sample set No. 3 are used as the initial state of agents, each agent takes 20 trajectories for Monte Carlo sampling; therefore, $n_\tau$ is 1,000. Since the agent takes 5 steps in each trajectory, there are 5,000 samples (1,000 trajectories times 5 steps) of $\{s_t, a_t, \hat{A}^{\pi_{\theta_k}}\}$ available for updating the actor and critic ANNs in that PPO iteration. The ANNs are trained by Adam [40] for 2,000 epochs in each PPO iteration. When using 200 airfoils in sample set No. 2, each agent takes 10 trajectories; therefore, $n_\tau$ is 2,000. The agents are realized in parallel computing to accelerate the sampling process. Then, the policy was updated with the trajectories and rewards from these multiple agents in one shot. The bias-variance tradeoff in GAE is set to 0.8, $\epsilon$ in PPO-clip is 0.1, and the discount ratio $\gamma$ is 0.99. The standard deviation of actions starts from 0.1.

After introducing the airfoil modification and evaluation methods, the environment can be defined. Algorithm C shows the process of an environment interacting with one agent, in which the state $s$ is $[X_1, M_{w,1}, M_{w,L}, M_{w,A}]$, the action $a$ is $[t_1, s_b, h_b]$, and the reward $r_k$ is $10,000 \times (C_{D,k} - C_{D,k+1})$. In a previous study [41], agents utilized the same baseline airfoil for training, and discrete actions were taken when interacting with the environment. In contrast, the present paper uses agents that take continuous actions starting from multiple baseline airfoils for reinforcement learning so that a more general policy can be obtained.

Algorithm C: algorithm of environments
___

1: Define the initial state $s_0$ of the agent, i.e., choose a baseline airfoil;
2: for k = 1, 2, …, $n_{max}$ (maximum steps)
3:    Collect the current state $s_k$, and the action $a_k$ agent takes based on its policy;
4:    Modify the airfoil using a bump function defined by $a_k$, and then reconstruct the airfoil;
5:    Calculate the next state $s_{k+1}$ and reward $r_k$ using CFD or surrogate models;
6:    If $M_{w,1} < 1.0$ (i.e., it is not a single shock wave airfoil), set the reward to zero and break;
7: end for
___

It should be noted that all the ANNs, including the surrogate models and the actor-critic models, are trained after the inputs and outputs are scaled to $(0,1)$. All the outputs are scaled back to real values before being passed to other modules. By scaling, the prediction error is easier to reduce, and the standard deviation of actions can use the same value for all components.

### III. Pretraining, PPO training and testing

Reinforcement learning can start from a blank slate, and good performances can be achieved under the right conditions. However, a randomly generated initial policy can be too hard to control, and it may never be improved. It is useful to pretrain the policy so that reinforcement learning can start from a robust and reasonable initial policy.

This section first pretrains the policy and value function. Then, the pretrained model is used for the initial parameters of the proximal policy optimization. The policy is improved by interacting with environments based on surrogate models, and the effect of pretraining is also discussed. Finally, the policy is tested by CFD under different flow conditions using the 35 typical airfoils from sample set No. 4.

### A. Pretraining of the policy and value function

Pretraining is an effective approach for improving reinforcement learning performance. There have been studies that pretrain the policy with expert demonstrations [42] or good trajectories [43] or pretrain other neural networks other than the policy [44]. Imitation learning is a supervised learning method that fits the policy ANN with demonstrations. Similar work [42] has shown that it would cause the policy to learn to imitate the "policy" that generates the demonstrative state-action samples.

In this paper, the state-action samples are obtained from greedy searches on surrogate models. Then, the policy ANN is updated by regression, after which the value function ANN is updated using a modified PPO-clip algorithm.

For each of the 200 airfoils in sample set No. 2, four searches with five steps are conducted to collect good state-action samples. The search algorithm is described by Algorithm D. The sampling produces 4,000 state-action samples. However, because four random searches are carried out from each baseline airfoil, there are many duplicated or conflicted samples. In other words, there are samples having different action and reward values for the same states. Therefore, only the sample with the highest reward is kept for samples that have the same states.

Algorithm D: searching algorithm for state-action samples

| | |
|---|---|
| 1: | Initialize the CST parameters $x_0$ of the baseline airfoil; |
| 2: | for k = 1, 2, …, 5 (steps); |
| 3: | Randomly generate 200 actions $a_{k,i}$ (i=1, 2, …, 200); |
| 4: | Get the new geometries $x_{k,i}$ by taking these actions to the current geometry $x_{k-1}$ |
| 5: | Evaluate the drag coefficient $C_{D,k,i}$, state $s_{k,i}$ and reward $r_{k,i}$ of $x_{k,i}$ based on surrogate models; |
| 5: | Select the action $a_{k,j}$ that achieves the smallest drag and satisfies the constraints in Eq. 4; |
| 6: | Update the geometry $x_k = x_{k,j}$, save $(s_{k,j}, a_{k,j}, r_{k,i})$ into the state-action sample set; |
| 7: | end for |

After the selection, 1,759 samples remained in the sample set. Due to the randomness of taking actions, the samples do not have a smooth distribution of actions. In other words, the action may change rapidly when the state is changed, which indicates a sensitive policy that cannot be used as the initial policy for reinforcement learning. Therefore, the samples need to be filtered to eliminate the high-frequency fluctuation in actions. The present paper utilizes a simple algorithm to smooth the data, which is described by Algorithm E.

Algorithm E: smoothing samples

1: Calculate Euclidean distances of states between all samples, $d_{i,j}$ ($i,j$=1, 2, ..., $n_{sample}$);
2: for k = 1, 2, ..., 10 (steps of smoothing);
3:   Calculate the average action of each sample using inverse distance weighted interpolation:
     First, find the nearest 10 samples, $j \in J$;
     Then, $\bar{a}_i = \sum_{j \in J}[(a_j/d_{i,j})/\sum_{l \in J}(1/d_{i,l})]$, $i$=1, 2, ..., $n_{sample}$;
4:   Update each sample by $a_i = a_i + (\bar{a}_i - a_i) \times 0.2$;
5: end for

The policy ANN is trained by regression using the selected and smoothed samples. The Adam algorithm is employed to minimize the loss function, i.e., $\text{loss}_{actor} = \sum_{i=1}^{n_{sample}}[\hat{a} - a_i]^2$, where $\hat{a} = \arg\max_{a} \pi_\theta(s, a)$. $\hat{a}$ is the mean value of the Gaussian distribution of stochastic policy action, which is the output layer of actor ANN in Figure 13. The training process is plotted in Figure 15. The initial learning rate is 0.001, and the learning rate is decayed by 0.1 once the number of epochs reaches 250, 500, and 750. The results show that the smoothed samples can produce a smaller regression loss. Furthermore, the state-action samples should be smoothed to avoid overfitting when deeper neural networks are used in future studies.

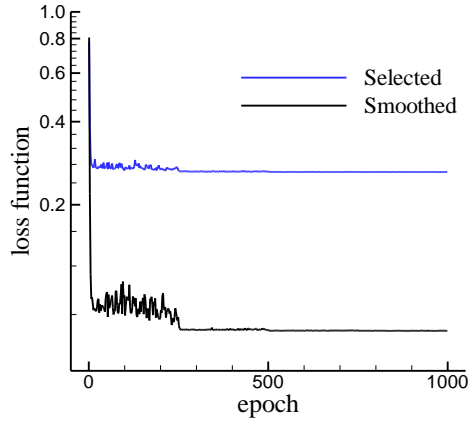

**Figure 15 Loss history of imitation learning**

After the imitation learning of the policy, the value function is updated by the PPO-clip algorithm without updating the actor., i.e., neglecting the 6$^{th}$ step so that the value function is learned to represent the current policy. The critic is updated using the 50 airfoils in sample set No. 3 as the initial state of agents. The learning rate is 0.01 for the first 10000 iterations and 0.001 for the next 10000 iterations. The training process is plotted in Figure 16.

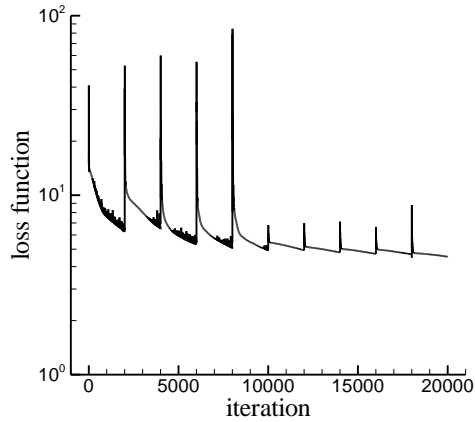

**Figure 16 Loss history of updating the critic model**

## B. Process of the proximal policy optimization

Because the proximal policy optimization algorithm is on-policy method, i.e., actions are sampled by its own policy, the trajectories for updating the policy are strongly dependent on its current behavior. When all the sampled actions cannot improve the policy via gradient descent, it is quite possible that the "policy optimization" will become stuck in local optima. Therefore, it is necessary to provide a reasonable and robust initial policy by pretraining so that the PPO algorithm is more likely to avoid local optima and learn a better policy.

In this section, both the pretrained policy and randomly generated policies are used to compare the training process and rewards. The policy is trained using the 50 airfoils in sample set No. 3 as the initial state of agents, and each airfoil generates 20 trajectories. As explained in Section II.E, there are 5,000 samples (50 baseline airfoils × 20 trajectories × 5 steps = 5,000 samples) of $\{s_t, a_t, \hat{A}^{\pi_{\theta_k}}\}$ available in each PPO iteration. The actor and critic ANNs are updated with these samples [40] for 2,000 epochs using Adam. The learning rate of the actor is $1 \times 10^{-6}$ for 200 PPO iterations, $1 \times 10^{-7}$ for the next 100 PPO iterations, and $1 \times 10^{-8}$ for another 100 PPO iterations. The learning rate for updating the critic is ten times that of the actor so that it can achieve a more robust training process in response to the large standard deviation of 0.1. The small learning rates ensure the policy not being updated too much in one PPO iteration, when the amount of training epochs is relatively large. The combination of small learning rates and large training epochs is found to be able to achieve a robust training process.

The pretrained policy, as well as six randomly generated initial policies, are used for PPO training. After updating the actor and critic in each PPO iteration, the policy is tested by taking the mean action, i.e., the standard deviation of actions is zero, and the cumulative rewards of the 50 airfoils are collected. Figure 17 shows the increase in the mean cumulative rewards of the 50 airfoils, i.e., the average drag coefficient reduction of these airfoils after the 5 steps of geometry modification. It can be seen that the pretrained policy (black line) can achieve the largest mean cumulative reward. In contrast, the randomly generated policies (color lines) do not guarantee good performance after training. Even though one of the randomly generated policies, i.e., the red line in Figure 17, achieves a similar mean cumulative reward with the pretrained policy, the final policy is not very satisfying. Figure 18 shows a typical airfoil modification process of that red line policy. Figure 18 (a) and (b) show the airfoil geometry change during the modification, and Figure 18 (c) shows the bump functions corresponding to the policy actions. The action parameters $[t_1, s_b, h_b]$, i.e., the parameters of bump functions, are shown in the legend.

Figure 18 (c) and Figure 19 (c) show that the policy takes different actions for different wall Mach number distributions. But the red line policy always reduces the leading-edge radius, whereas the pretrained policy is more flexible. Additionally, the red line policy tends to compromise the low-speed performances and the buffet characteristics by reducing the leading-edge radius. Therefore, it is preferable to provide reinforcement learning with a reasonable initial policy by pretraining.

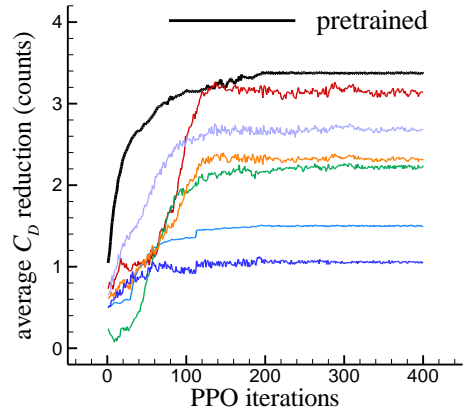

**Figure 17 Average drag reduction of 50 airfoils during PPO training**

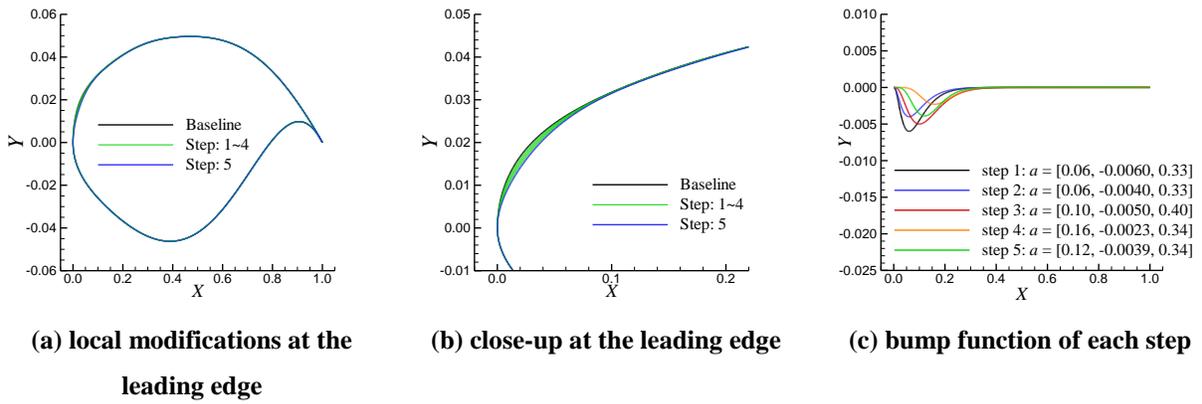

(a) local modifications at the leading edge

(b) close-up at the leading edge

(c) bump function of each step

**Figure 18 Typical airfoil modification process of the trained policy (random)**

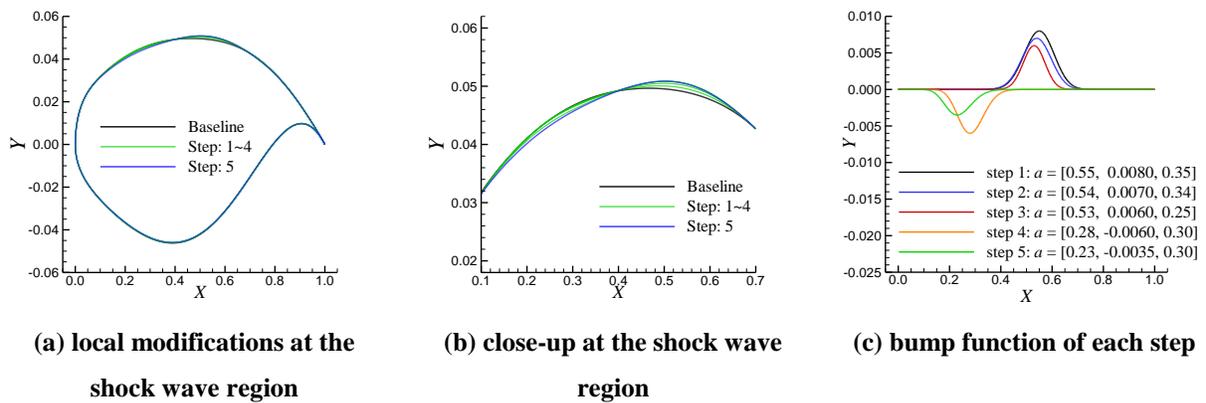

(a) local modifications at the shock wave region

(b) close-up at the shock wave region

(c) bump function of each step

**Figure 19 Typical airfoil modification process of the trained policy (pretrained)**

In the following discussions, the policy, by which the initial policy is the pretrained policy in section III.A, is further trained and tested. After training with 50 airfoils, the policy is further trained using the 200 airfoils in sample set No. 2 as the initial agent states. The initial policy, i.e., the pretrained policy achieves an average drag reduction of 1.34 drag counts when tested by the 200 airfoils. The average drag reduction of 200 airfoils is increased to 3.71 after being trained with 50 airfoils, which is the PPO training process shown in Figure 17. Then, the policy is further trained with the 200 airfoils with the PPO algorithm. The learning rate of the actor is $1 \times 10^{-7}$ for 200 PPO iterations, $1 \times 10^{-8}$ for the next 200 PPO iterations, and $1 \times 10^{-9}$ for another 200 PPO iterations. The final average drag reduction is 5.53 drag counts. Figure 20 shows the history of average drag reduction during PPO training by 200 airfoils, which proves that the policy can be further improved when more baseline airfoils are used.

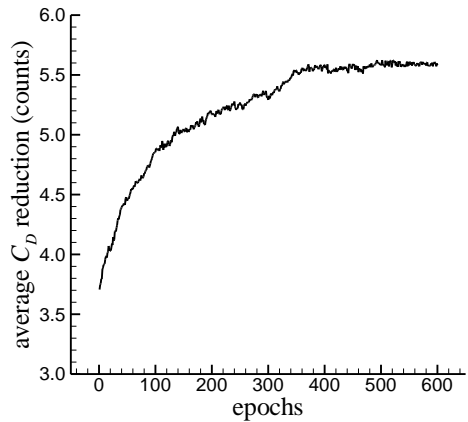

**Figure 20 Average drag reduction of 200 airfoils**

After reinforcement learning, both the initial policy and trained policies are tested by 35 test airfoils from sample set No. 4. The drag reductions of the 35 airfoils during the different phases of PPO training are listed in Table 1. The results show that the average drag reduction of 35 airfoils is 1.02 drag counts when the initial policy is employed, and then the average drag reduction is 2.66 drag counts after PPO training with 50 airfoils. It is further increased to 3.17 drag counts after training with 200 airfoils. This indicates that the policy is valid in the testing set as well.

**Table 1 Drag reductions of the 35 airfoils**

|   | Drag reduction (counts) | Description |
|---|---|---|
| 1 | 1.02 | Initial policy |
| 2 | 2.66 | PPO training with 50 airfoils |
| 3 | 3.17 | PPO training with 200 airfoils |

## C. Policy tests by CFD

Reinforcement learning has been proven to significantly improve the policy performance on surrogate models when the policy is tested by the training airfoils in the training condition, i.e., $M_\infty$=0.76, $C_L$=0.70, $Re$=5 × 10$^6$, and $t_{max}$=0.095. This section tests the drag reduction ability when the policy is used in CFD environments. In this section, policy is tested by 35 typical airfoils from sample set No. 4. Five steps are taken for each airfoil. During the test, the policy takes the mean value of its action output, i.e., the standard deviation of actions is zero.

The policy is tested in the training condition so that a baseline of policy performance in CFD environments is provided, its average drag reduction is 2.77 counts. Although the rewards in the CFD environments do not exactly match the results on surrogate models, the policy trained on surrogate models can still achieve drag reductions. The policy performance in CFD environments improves when more accurate surrogate models are utilized. Figure 21 shows the wall Mach number distributions during the modification processes of two test airfoils (No. 2 and No. 31) in sample set No. 4. Figure 22 is the modification processes of the airfoil geometry. Since the pretraining in section III.A used selected action samples that act around the shock wave, the pretrained policy only modifies the shock wave region. Consequently, the learned policy mostly focuses on the shock wave region, as shown in the two demonstrations in Figure 22. Figure 23 shows the drag coefficient history during the modification, the results show that the policy can effectively weaken the shock wave, and the drag coefficient can also be reduced in several steps. Therefore, reinforcement learning can learn a general policy that works for different airfoils in CFD environments.

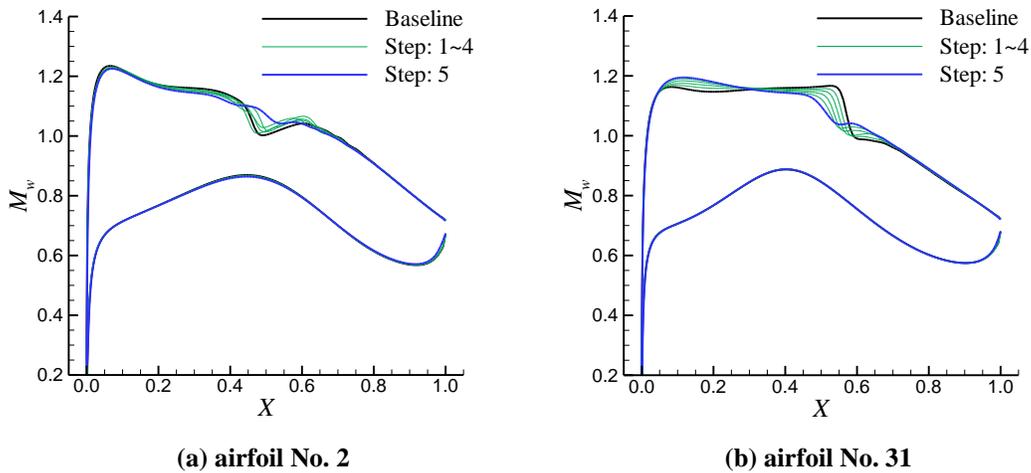

(a) airfoil No. 2  (b) airfoil No. 31

**Figure 21 Wall Mach number distributions during airfoil modification (training condition)**

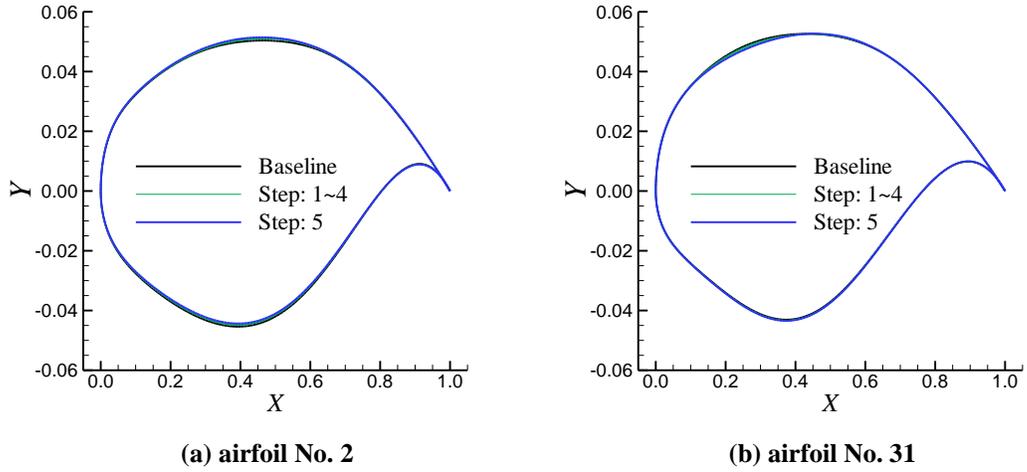

**(a) airfoil No. 2**  **(b) airfoil No. 31**

**Figure 22 Airfoil geometries during airfoil modification (training condition)**

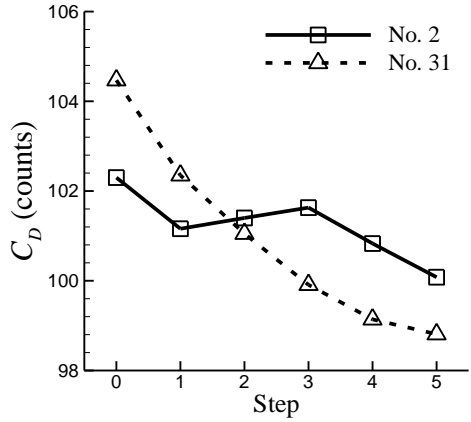

**Figure 23 History of drag coefficients (training condition)**

### D. Test of transfer applications by CFD

The present paper uses features of airfoil wall Mach number distributions as the state parameters so that the learned policies can be attributed to the general laws of aerodynamics. Theoretically, reinforcement learning can obtain a general policy that can be applied to any supercritical airfoil in similar transonic flow conditions, without the need of further training in the new condition.

The learned policy is first tested by CFD in six test conditions, which have different free-stream Mach numbers, lift coefficients, Reynolds numbers, and airfoil maximum thicknesses. The test conditions are listed in Table 2, in which the 6[th] condition is completely different from the training condition. The 35 airfoils in sample set No. 4 are used for baselines, and the policy takes 5 steps of modification to each airfoil. The average drag reduction of 35 airfoils are also listed in Table 2. The results show that the learned policy can effectively reduce the drag of different airfoils in different flow conditions. In the six test cases, the learned policy based on physical features can achieve 8.64 counts of drag reduction on average.

**Table 2 Test conditions and average drag reduction of 35 test airfoils (counts)**

| Case | $M_\infty$ | $C_L$ | Re | $t_{max}$ | learned policy (5 steps) |
|---|---|---|---|---|---|
| 1 | 0.76 | 0.70 | $8 \times 10^6$ | 0.095 | 3.30 |
| 2 | 0.77 | 0.70 | $5 \times 10^6$ | 0.095 | 9.91 |
| 3 | 0.76 | 0.80 | $5 \times 10^6$ | 0.095 | 7.69 |
| 4 | 0.76 | 0.70 | $5 \times 10^6$ | 0.105 | 10.90 |
| 5 | 0.75 | 0.80 | $8 \times 10^6$ | 0.105 | 8.81 |
| 6 | 0.72 | 0.85 | $1 \times 10^7$ | 0.135 | 11.25 |

To test the ability of the policy taking more steps, the learned policy is employed to take 20 steps of modifications. The policy has taken 5 steps of modification in the previous tests, of which the results are shown in Table 2. The policy is employed to take 15 more steps to achieve the maximum drag reduction for each airfoil. Figure 24 shows the drag reduction of the learned policy in the 6$^{th}$ test condition, when 5 steps and 20 steps of modification are taken. The black line with scatters (, i.e., group "baseline airfoil") shows the drag coefficients of 35 baseline airfoils in this test condition, in which the airfoils are sorted in order of drag coefficient. The black bars with gray shade (, i.e., group "final airfoil") are the drag coefficients of airfoils after 20 steps of modification. Then, the height of dark blue bars (, i.e., group "drag reduction in 5 steps") shows the drag reduction during the first five modification steps, and the height of light blue bars (, i.e., group "drag reduction in another 15 steps") shows the drag reduction during the remaining 15 steps.

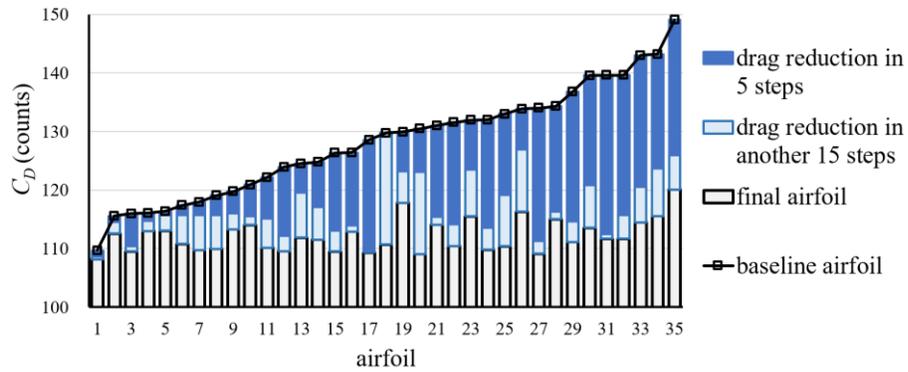

**Figure 24 Drag reductions of 35 airfoils with 5 and 20 modification steps (6$^{th}$ test condition)**

Figure 24 shows that the learned policy can further reduce drag when more steps are taken, and the learned policy can eventually achieve low drag coefficients for all test airfoils. Figure 25 shows the drag reduction history of test airfoil No. 2 and No. 31. (The two test airfoils are the same two airfoils in Figure 21, but they are the 4$^{th}$ airfoil and 27$^{th}$ airfoil in Figure 24, because the airfoils in Figure 24 are sorted in the order of drag coefficients in the test condition.) It can be seen from Figure 24 and Figure 25 that the most portion of drag reduction is achieved in the first five steps, which indicates the high efficiency of policy and the good performance of reinforcement learning.

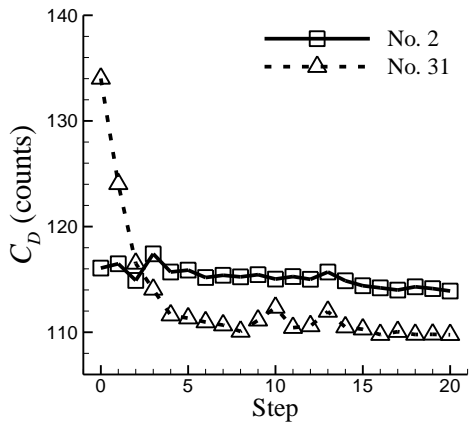

**Figure 25 History of drag coefficients (6$^{th}$ test condition)**

Figure 26 shows the wall Mach number distributions during the modifications of airfoil No. 2 and No. 31. Although the policy does not achieve shockless designs, airfoils with very weak shock waves are obtained. The results show that, for the final modification results of these two airfoils, the angle of attack equals 1.51 and 0.98 degrees, respectively. Therefore, although both airfoils have similar weak shock waves, the final drag coefficient of airfoil No. 31 is smaller than airfoil No. 2, as shown in Figure 25.

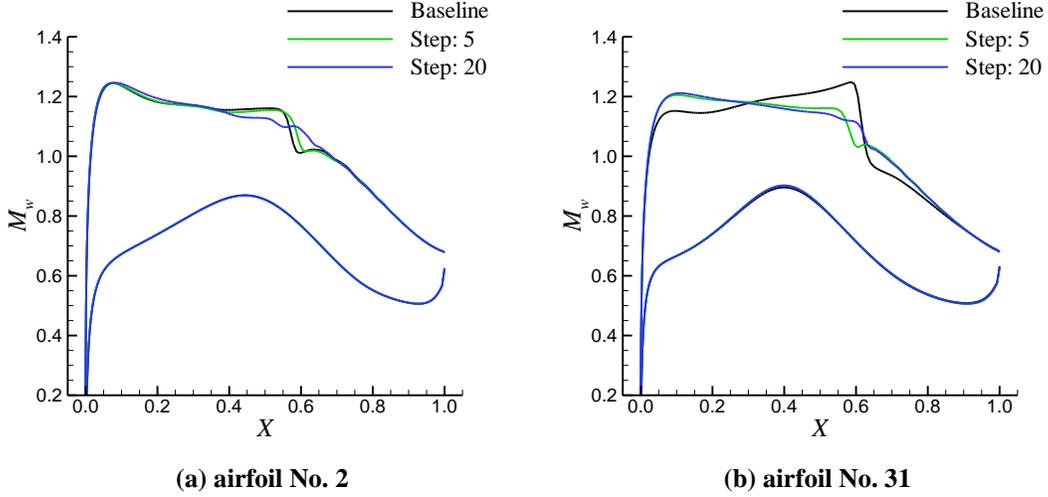

**(a) airfoil No. 2**  **(b) airfoil No. 31**

**Figure 26 Wall Mach number distributions during airfoil modification (6$^{th}$ test condition)**

To further demonstrate the transfer application ability, the comparison between the learned policy, surrogate-based optimizations and genetic optimizations is conducted. The learned policy is employed in the CFD environment to take 20 steps of modification in the training condition (case 0) and six test conditions (case 1~6), so that the largest drag reduction by the learned policy is achieved. The average drag reduction of 35 airfoils are listed in Table 3.

Surrogate-based optimizations are conducted to demonstrate the limit of transfer application for geometry-based surrogate models or policies. Physical features are used for reinforcement learning in this paper, so that better transfer application ability can be achieved. In contrast, policies based on airfoil geometries are not aware of the change in flow condition since their state only has geometric parameters. Strictly speaking, geometry-based policies can only be applied for the training condition. Similarly, surrogate-based optimizations [1] that use geometric parameters as surrogate inputs also have limited transfer application ability for the same reason. Theoretically, surrogate-based optimizations need to build new surrogate models for new flow conditions, in other words, the surrogate model built in the training condition cannot be transferred to test cases. In this section, the surrogate model built in Section II.D is used for transfer applications in the six test cases, and the 'optimized' geometries of 35 test airfoils are obtained by optimizations on that surrogate model. The actual drag reduction of the 'optimized' airfoils are validated by CFD in the training condition and six test conditions, the results are listed in Table 3.

The optimizations utilize a genetic algorithm from an open source software jMetalpy, which employs the simulated binary crossover, polynomial mutation and tournament selection [45]. The optimization problem is described in Eq. 6. The independent variables are the seven CST parameters of the airfoil upper surface, and the airfoil maximum thickness is kept the same during optimization. The above setup is to keep the optimization problem the same as the modification process by the learned policy. It means 35 optimizations need to be carried out for the 35 test airfoils, respectively, because they have different lower surface geometries. The population size is 25, total optimization generation is 40, and the initial population is generated by random perturbation to the baseline airfoil.

$$\min C_D$$
$$\text{s.t. single shock wave} \tag{6}$$

Genetic optimizations are also conducted with CFD to find the 'actual' largest drag reductions, which can be used as reference values to evaluate the performance of the learned policy. The genetic optimizations have the same setup as the surrogate-based optimizations, except that they are conducted with CFD instead on surrogate models. The drag reductions in the training condition and six test conditions are also listed in Table 3.

**Table 3 Average drag reductions of 35 test airfoils (counts)**

| Case | learned policy (20 steps) | surrogate-based optimization | genetic optimization |
|---|---|---|---|
| 0 | 2.79 | 2.78 | 2.81 |
| 1 | 4.11 | 2.19 | 4.38 |
| 2 | 11.57 | 4.51 | 12.41 |
| 3 | 12.21 | 4.86 | 14.58 |
| 4 | 13.55 | 3.34 | 14.24 |
| 5 | 12.18 | 5.39 | 13.24 |
| 6 | 16.08 | 3.45 | 18.59 |
| CFD evaluations | 21 | 1 | 1,000 |

Table 3 shows that the learned policy can achieve similar performances with genetic optimizations, both in the training condition and transfer applications. In contrast, although the surrogate-based optimization also achieves a similar performance in the training condition, it has relatively poor performances in transfer applications. Additionally, the last row of Table 3 shows the amount of CFD evaluations for obtaining the 'optimized' airfoil by different methods. The surrogate-based optimization only needs one CFD evaluation to validation the result predicted by the surrogate model. The genetic optimization takes 1,000 CFD evaluations to converge for optimizing each baseline airfoil in each test case, and the learned policy takes 21 CFD evaluations to obtain similar results. Although it may take hundreds of thousands of CFD evaluations to train a policy, the policy is theoretically applicable for other similar cases. Therefore, the training costs can be shared by many applications, which is the benefit of a good transfer application ability.

# IV. Conclusions

Designers modify airfoils using their rich and general experiences, which are based on observations of performances, geometries and physical features of the flow field. Reinforcement learning is an artificial general intelligence technology that can learn this kind of experience by a vast number of trial-and-errors. The present paper utilized the proximal policy optimization algorithm to learn the policy of reducing the drag of supercritical airfoils, before which the policy was pretrained through imitation learning to provide a good initial policy. Then, the policy was tested on both surrogate models and CFD environments. The following conclusions have been reached.

(1) Since reinforcement learning requires a large number of performance evaluations during training, surrogate models were used as rapid analysis methods to reduce computational costs. In the drag reduction of supercritical airfoils, reinforcement learning increased the mean cumulative reward, i.e., average drag reduction, of 200 airfoils from 1.34 to 5.53. The results showed that the policy trained on surrogate models was still effective in CFD environments.

(2) Randomly generated policies can sometimes be difficult for reinforcement learning to improve. In contrast, a robust and reasonable initial policy can be obtained through imitation learning, which effectively improves the performance of reinforcement learning. It was also recommended to smooth the state-action samples for imitation learning to avoid overfitting the policy in regression.

(3) To improve the ability of transfer application, the policies preferably include physical features in the state parameters, and wall Mach number distribution features were used in the present paper. This is because the relationship between aerodynamic performances and wall Mach number distribution features is more general for different airfoils and flow conditions so that the policy can be more general when it makes decisions based on physical features. The results showed that in transfer applications, the policy based on wall Mach number distribution features could still effectively reduce drag of various supercritical airfoils in different flow conditions, and more steps could be taken to further reduce drag.

In summary, reinforcement learning can learn the policies of airfoil aerodynamic designs. The performance of reinforcement learning can be improved when the initial policy is pretrained by imitation learning, and the policy can achieve the better transfer application ability when physical features are used for state parameters.

## Acknowledgments

This work was supported by the National Natural Science Foundation of China under Grant Nos. 91852108 and 11872230.

## References

[1] Queipo, N. V., Haftka, R. T., Shyy, W., and Tucker, P. K., "Surrogate-based analysis and optimization," Progress in Aerospace Sciences, vol.41, 2005, pp. 1-28.
doi: https://doi.org/10.1016/j.paerosci.2005.02.001

[2] Han, Z.H., Zhang, Y., and Zhang, K.S, "Weighted gradient-enhanced kriging for high-dimensional surrogate modeling and design optimization," AIAA Journal, vol.55, No.12, 2017, pp. 4330–4346.
doi: https://doi.org/10.2514/1.J055842

[3] Soilahoudine, M., Gogu, C., and Bes, C., "Accelerated adaptive surrogate-based optimization through reduced-order modeling," AIAA Journal, vol.55, No.5, 2017, pp. 1681-1694.
doi: https://doi.org/10.2541/1.J055252

[4] Li, R., Deng, K., and Chen, H., "Pressure distribution guided supercritical wing optimization," Chinese Journal of Aeronautics, vol.31, 2018, pp.1842-1854.
doi: https://doi.org/10.1016/j.cja.2018.06.021

[5] Sun, G., and Wang, S., "A review of the artificial neural network surrogate modeling in aerodynamic design," Journal of Aerospace Engineering, vol.233, 2019, pp. 5863 - 5872.
doi: https://doi.org/10.1177/0954410019864485

[6] Hu, L., Zhang, J., and Wang, W., "Neural networks-based aerodynamic data modeling: a comprehensive review," IEEE Access, vol.8, 2020, pp. 90805–90823.
doi: https://doi.org/10.1109/ACCESS.2020.2993562

[7] Cao, Z., Xu, J., Xiao, and Wu, H., "A novel method for detection of wind turbine blade imbalance based on multi-variable spectrum imaging and convolutional neural network," 2019 Chinese Control Conference, 2019, pp. 4925-4930.

[8] Chen, H., He, L., and Wang, S., "Multiple aerodynamic coefficient prediction of airfoils using a convolutional neural network," Symmetry, vol. 12, 2020, 544.
doi: https://doi.org/10.3390/sym12040544

[9] Chen, W., Chiu, K., and Fuge, M. D., "Aerodynamic design optimization and shape exploration using generative adversarial networks," AIAA Scitech Forum, 2019.
doi: https://doi.org/10.2514/6.2019-2351

[10] Du, X., He, P., and Martins, J., "A B-spline-based generative adversarial network model for fast interactive airfoil aerodynamic optimization," AIAA Scitech Forum, 2020.
doi: https://doi.org/10.2514/6.2020-2128

[11] Chua, J. C., Lopez, N. S., and Augusto, G.., "Numerical and statistical analyses of aerodynamic characteristics of low Reynolds number airfoils using Xfoil and JMP," AIP Conference Proceedings, vol. 1905, 2017, pp. 50016.
doi: https://doi.org/10.1063/1.5012235

[12] Li, J., Mohamed, A., and Martins, J., "Data-based approach for fast airfoil analysis and optimization," AIAA Journal, vol.57, No.2, 2019, pp. 581-596.
doi: https://doi.org/10.2514/1.J057129

[13] Li, R., Zhang, Y. and Chen, H., "Design of Experiment Method in Objective Space for Machine Learning of Flow Structures," 8th European Conference for Aeronautics and Space Sciences, Madrid, Spain, 2019.
doi: https://doi.org/10.13009/EUCASS2019-403

[14] Sutton, R., and Barto, A., "Reinforcement Learning: An Introduction," IEEE Transactions on Neural Networks, vol. 16, 2005, pp. 285-286.
doi: https://doi.org/10.1109/TNN.1998.712192

[15] Li, Y., "Deep Reinforcement Learning: An Overview," ArXiv Preprint, 2017, 1701.07274.

[16] Mnih, V., Kavukcuoglu, K., and Ostrovski, G., (2015). Human-level control through deep reinforcement learning. Nature, vol. 518, No.7540, 2015, pp. 529–533.
doi: https://doi.org/10.1038/NATURE14236

[17] Silver, D., Huang, A., and Lanctot, M., "Mastering the game of Go with deep neural networks and tree search," Nature, vol. 529, No.7587, 2016, pp. 484–489.
doi: https://doi.org/10.1038/NATURE16961

[18] Kober, J., Bagnell, J. A., and Peters, J., "Reinforcement learning in robotics: A survey," The International Journal of Robotics Research, vol. 32, No.11, 2013, pp. 1238–1274.
doi: https://doi.org/10.1177/0278364913495721

[19] Mirhoseini, A., Goldie, A., and Bae, S., "Chip Placement with Deep Reinforcement Learning," ArXiv Preprint, 2020, 2004.10746.

[20] Garnier, P., Viquerat, J., and Hachem, E., "A review on deep reinforcement learning for fluid mechanics," ArXiv Preprint, 2019, 1908.04127.

[21] Rabault, J., Ren, F., and Xu, H., "Deep reinforcement learning in fluid mechanics: A promising method for both active flow control and shape optimization," Journal of Hydrodynamics, vol. 32, No.2, 2020, pp. 234–246.
doi: https://doi.org/10.1007/S42241-020-0028-Y

[22] Yan, X., Zhu, J., and Wang, X, "Missile aerodynamic design using reinforcement learning and transfer learning," Science in China Series F: Information Sciences, vol. 61, No.11, 2018, pp. 119204.
doi: https://doi.org/10.1007/S11432-018-9463-X

[23] Viquerat, J., Rabault, J., and Hachem, E, "Direct shape optimization through deep reinforcement learning," ArXiv Preprint, 2019, 1908.09885.

[24] Polydoros, A. S., and Nalpantidis, L., "Survey of model-based reinforcement learning: applications on robotics," Journal of Intelligent and Robotic Systems, vol. 86, No.2, 2017, pp. 153–173.

doi: https://doi.org/10.1007/S10846-017-0468-Y

[25] Haarnoja, T., Zhou, A., and Levine, S., "Soft actor-critic: off-policy maximum entropy deep reinforcement learning with a stochastic actor," International Conference on Learning Representations, 2018.

[26] Viquerat, J, and Elie H., "A supervised neural network for drag prediction of arbitrary 2D shapes in laminar flows at low Reynolds number," Computers & Fluids, vol. 210, 2020, pp. 104645.
doi: https://doi.org/10.1016/J.COMPFLUID.2020.104645

[27] Ma, P., Tian, Y., and Monacha, D., "Fluid directed rigid body control using deep reinforcement learning," ACM Transactions on Graphics, vol. 37, No. 4, 2018, pp. 96.
doi: https://doi.org/10.1145/3197517.3201334

[28] Li, R., Zhang, Y. and Chen, H., Strategies and methods for multi-objective aerodynamic optimization design for supercritical wings[J]. Acta Aeronautica et Astronautica Sinica, vol.41, 2020, pp.623409.
doi: https://doi.org/0.7527/S1000-6893.2019.23409

[29] Délery, J., Marvin, J.G. and Reshotko, E., "Shock-wave boundary layer interactions," Advisory Group for Aerospace Research and Development Neuilly-Sur-Seine (France), 1986, No. AGARD-AG-280.

[30] Kulfan, B. M., "Universal parametric geometry representation method,"Journal of Aircraft, vol. 45, No. 1, 2008, pp. 142-158.
doi: https://doi.org/10.2514/1.29958

[31] Castonguay, P., and Nadarajah, S., "Effect of Shape Parameterization on Aerodynamic Shape Optimization," 45th AIAA Aerospace Sciences Meeting and Exhibit, Nevada, 2007.
doi: https://doi.org/10.2514/6.2007-59

[32] https://cfl3d.larc.nasa.gov/

[33] Cook, P. H., Mcdonald, M. A., Firmin, M. C. P. "Aerofoil RAE 2822—pressure distributions, and boundary layer and wake measurements," Experimental Data Base for Computer Program Assessment, AGARD Advisory Report No. 138. pp. A6-1~76, May 1979.

[34] Li, R., Zhang, Y. and Chen, H., "Design of Experiment Method in Objective Space for Machine Learning of Flow Structures," 8th European Conference for Aeronautics and Space Sciences, Madrid, Spain, 2019.
doi: https://doi.org/10.13009/EUCASS2019-403

[35] Steiner, B., DeVito, Z., and Yang, E., "PyTorch: an imperative style, high-performance deep learning library," 33th Conference on Neural Information Processing Systems, 2019, pp. 8026–8037.

[36] Cotter, A., Shamir, O., and Sridharan, K., "Better mini-batch algorithms via accelerated gradient methods," Advances in Neural Information Processing Systems, 2011, pp. 1647–1655.

[37] Schulman, J., Wolski, F., and Klimov, O., "Proximal policy optimization algorithms," ArXiv Preprint, 2017, 1707.06347.

[38] Schulman, J., Levine, S., and Moritz, P., "Trust region policy optimization," Proceedings of the 32nd International Conference on Machine Learning, 2015, pp. 1889–1897.

[39] Schulman, J., Moritz, P., and Abbeel, P., "High-dimensional continuous control using generalized advantage estimation," International Conference on Learning Representations, 2016.

[40] Kingma, D. P., and Ba, J. L., "Adam: a method for stochastic optimization," International Conference on Learning Representations, 2015.

[41] Li, R., Zhang, Y., and Chen, H., "Study of reinforcement learning method for supercritical airfoil aerodynamic design," Acta Aeronauticaet Astronautica Sinica, 2020, vol.45, 23810.
doi: https://doi.org/10.7527/S1000-6893.2020.23810

[42] Zhang, X., Ma, H., "Pretraining deep actor-critic reinforcement learning algorithms with expert demonstrations," ArXiv Preprint, 2018, 1801.10459.

[43] Ross, S., Gordon, G. J., and Bagnell, J. A., "A reduction of imitation learning and structured prediction to no-regret online learning," Proceedings of the Fourteenth International Conference on Artificial Intelligence and Statistics, vol. 15, 2011, pp. 627–635.

[44] Cruz, G. V., Du, Y., and Taylor, M. E., "Pre-training with non-expert human demonstration for deep reinforcement learning," Knowledge Engineering Review, vol. 34, 2019.
doi: https://doi.org/10.1017/S0269888919000055

[45] Benítez-Hidalgo, A., Nebro, A. J., García-Nieto, J., Oregi, I., and Ser, J. D., "jMetalPy: A Python framework for multi-objective optimization with metaheuristics," Swarm and Evolutionary Computation, vol. 51, 2019, 100598.
doi: https://doi.org/10.1016/J.SWEVO.2019.100598